# COPING WITH INDUCTIVE RISK WHEN THEORIES ARE UNDERDETERMINED: DECISION MAKING WITH PARTIAL IDENTIFICATION

Charles F. Manski

Department of Economics and Institute for Policy Research
Northwestern University, Evanston, IL 60208 USA
cfmanski@northwestern.edu
ORCID: 0000-0001-7260-7686

Abstract

Controversy about the significance of underdetermination of theories persists in the philosophy and conduct of science. The issue has practical import when research is used to inform decision making, because scientific uncertainty yields inductive risk. Seeking to enhance communication between philosophers and researchers who study public policy, this paper describes econometric analysis of *partial identification* and its use in welfare-economic policy analysis. Study of partial identification finds underdetermination and inductive risk to be highly consequential for credible prediction of important societal outcomes and, hence, for credible public decision making. It provides mathematical tools to characterize a broad class of scientific uncertainties that arise when available data and well-supported assumptions are combined to predict population outcomes. Combining study of partial identification with criteria for reasonable decision making under uncertainty yields coherent approaches to make policy choices without accepting one among multiple empirically underdetermined theories. The paper argues that study of partial identification warrants attention in philosophical discourse on underdetermination and inductive risk.

## 1. Introduction

In *A System of Logic*, John Stuart Mill cogently called attention to the empirical underdetermination of theories, writing (Mill, Eighth Edition, 1881, p. 356):

> "Accordingly, most thinkers of any degree of sobriety allow that an hypothesis of this kind is not to be received as probably true because it accounts for all the known phenomena; since this is a condition sometimes fulfilled tolerably well by two conflicting hypotheses; while there are probably many others which are equally possible, but which, for want of any thing analogous in our experience, our minds are unfitted to conceive."

Mill was a notable precursor to the modern concern with underdetermination in the philosophy of science.

In a recent entry in the *Stanford Encyclopedia of Philosophy*, Stanford (2023) discussed the divergence of views between philosophers who consider underdetermination to pose a fundamental problem for science and those who view it as a minor concern. Stanford judged the latter perspective to be more prevalent currently, writing (p. 28): "most thinkers do not see why we should believe the least we can get away with rather than believing the most we are entitled to by the evidence we have." However, he concluded that the question remains open, writing (p. 32-33):

> "Thus, claims and arguments about the various forms that underdetermination may take, their causes and consequences, and the further significance they hold for the scientific enterprise as a whole continue to evolve in the light of ongoing controversy, and the underdetermination of scientific theory by evidence remains very much a live and unresolved issue in the philosophy of science."

The discussion in Turnbull (2017) is revealing as well. Turnbull cites the disparaging remark of Kitcher (2001, p.36) that underdetermination is merely a "philosopher's problem." However, she does not concur. She writes (p. 7).

> "Unfortunately, discussions of underdetermination have often framed underdetermination as a 'philosopher's problem' rather than as a scientist's problem . . . But underdetermination is not solely a philosopher's problem. Practical Underdetermination occurs frequently in the practice of science and is easily recognizable by scientists themselves."

Controversy about the significance of underdetermination is a live and unresolved issue not only in philosophy but also in the conduct of science. The issue has practical import when scientific research is used to inform decision making, as scientific uncertainty yields *inductive risk* (Hempel, 1965). Douglas (2000) characterizes inductive risk as (p. 560): "the chance that one will be wrong in accepting (or rejecting) a scientific hypothesis."

The methodology used to define and quantify scientific error may perhaps be only an esoteric subject under the venerable perspective that science should be a "value-free" enterprise that operates independent of society. Summarizing that perspective, but not agreeing with it, Douglas (2000) wrote (p. 559): "With some exceptions . . . , the common wisdom of philosophers of science has been that only epistemic values have a legitimate role to play in science." The prevalence in philosophy of the view that science should be value-free appears to have diminished since 2000. Holman and Wilholt (2022) proclaim (p. 211):

> "The debate over whether non-epistemic values can play a legitimate role in science has largely come to a close, at least insofar as there is a growing acceptance amongst philosophers of science that values are an inherent part in the core processes of scientific inquiry."

Whether or not the debate has in fact ended, it is evident that consideration of values is inescapable when scientific uncertainty limits the ability of society to evaluate the impact of alternative public policies. Philosophers have recognized that coherent social planning under uncertainty requires attention to both knowledge and values (Rudner, 1953; Douglas, 2000; Holman and Wilmot, 2022).

### 1.1. Welfare Economics

Within the sciences, attention to the role of values in motivating research has been particularly explicit in welfare economics. Beginning informally in the late 1700s and increasingly formalized throughout the 1900s, economists have used the construct of a social welfare function (SWF) to compare public policies. Economists seek to characterize the social welfare achieved by alternative policies, aiming to find one that maximizes an SWF.

Economists studying policy choice in democracies strive to specify SWFs that in some manner express the values of population members rather than the preferences of dictators. Economists often study planning with utilitarian welfare functions. A utilitarian welfare function specifies interpersonally comparable cardinal personal welfares and sums them. This formalizes what Bentham (1776) may have had in mind when he wrote (p. ii): a "fundamental axiom, it is the greatest happiness of the greatest number that is the measure of right and wrong." They sometimes specify SWFs that express a form of paternalism or principles of fairness.

Paul Samuelson placed responsibility for specification of the SWF on society rather than on the economist, writing (Samuelson, 1947, p. 220): "It is a legitimate exercise of economic analysis to examine the consequences of various value judgments, whether or not they are shared by the theorist." Manski (2024) observed that economic studies of realistic classes of planning problems have commonly specified 'pragmatic' welfare functions, where 'pragmatic' means that (p. 31): "researchers motivate their welfare functions by some combination of conjecture regarding societal values, empirical study of population preferences, and concern for analytical tractability."

Although welfare economics has always recognized the centrality of values to policy formation, until recently it largely abstracted from incompleteness of knowledge of the impacts of alternative policies. In philosophical terms, it has abstracted from the fact that scientific uncertainty yields inductive risk. Whether performing abstract theoretical studies or applied benefit-cost analyses, economists have commonly assumed that the planner knows enough about the choice environment to be able to determine an optimal action.[1]

---

[1] The longstanding dearth of study of planning under uncertainty is apparent in the comprehensive textbook on public economics of Atkinson and Stiglitz (1980), which mentions uncertainty only a few times and then only in passing. Mongin and Pivato (2016) began their review article with this sentence (p. 711): "PERHAPS surprisingly, uncertainty plays no role whatsoever in the classical works on social welfare."

In reality, the consequences of policy decisions are often highly uncertain. Addressing the failure of welfare economics to come to grips with scientific uncertainty has motivated the development over the past twenty-five years of a program of welfare-economic research on social planning under uncertainty. In a recent monograph, Manski (2024) emphasizes that welfare-economic comparison of policies requires answers to several questions (p. 4): "What constitutes social welfare? What are the feasible actions? What is known about the welfare consequences of alternative choices?"

In the welfare-economic paradigm, values determine social welfare. The political, economic, and physical environment determine the feasible actions. The state of science determines the available knowledge of the consequence of alternative policy choices.

Viewed broadly, welfare economics considers the interaction of values and science in a manner similar to some classical social scientists and modern philosophers of science. Holman and Witholt (2022) interpret Max Weber as espousing the view that (p. 214):

> "What science was capable of yielding was an assessment of what means could achieve a predetermined set of goals and what the consequences of taking certain actions would be, but science did not determine what goals and principles should be pursued."

This statement is consistent with the perspective on welfare economics expressed by Samuelson (1947), but it is not equivalent. Weber worked in the late 1800s and early 1900s, before welfare economics became formalized mathematically and quantified numerically. The recent article by Holman and Witholt is written in the qualitative style prevalent in part of philosophy of science, with no mention of welfare economics. In contrast, from the mid-twentieth century onward, theoretical research in welfare economics has been mathematical and applied research has been quantitative.

These distinctions are not just details; they are fundamental. Economists find that policy choice seeking to maximize social welfare requires careful attention to tradeoffs. Policy A may be more advantageous than policy B in some respects, but less advantageous in others. Qualitative discussion is incapable of evaluating tradeoffs rigorously. Economists find that mathematical theory and numerical quantification are essential.

An apt example of tradeoffs occurs in medical treatment, when policy A is surveillance of a patient who is at risk of illness and B is aggressive treatment. Aggressive treatment is appealing to the extent that it better prevents disease onset or reduces the severity of illness. Surveillance is attractive to the extent that it avoids side effects that may occur with aggressive treatment. Making the optimal choice between surveillance and aggressive treatment requires knowledge of the relative magnitudes of the health effects of illness and side effects. Obtaining this knowledge requires quantitative analysis of the tradeoffs, not just qualitative recognition of their existence. See Manski (2019).

Another apt example of tradeoff occurs in formation of climate policy, when policy A takes moderate steps to reduce greenhouse gas (GHG) emissions and B seeks to aggressively reduce emissions. Economists studying climate policy consider a planner who seeks to make optimal trade-offs between the costs of GHG abatement and the economic damages from climate change. They use i*ntegrated assessment* (IA) models to evaluate alternative climate policies. These quantified mathematical models represent the climate and the links between the climatic effects of emissions and their impacts on the economy; see Nordhaus (2019). IA models make plain the fundamental difficulty in choosing between moderate and aggressive policies to reduce emissions. Moderate (aggressive) polices have lower (higher) abatement costs. Moderation yields greater (less) global warming and hence greater (smaller) potential damage from climate change. Appropriately quantifying these opposing tendencies is essential to optimal policy choice.

### 1.2. Partial Identification

Seeking to enhance communication between philosophers of science and economists who analyze public policy, this paper describes econometric analysis of *partial identification* and its application to policy analysis. This methodological research, whose formal structure will be described in Section 2, takes underdetermination of theories and inductive error very seriously, while using different terminology for these concepts. Study of partial identification finds underdetermination to be highly consequential for credible prediction of important societal outcomes and for credible formation of public policy. It provides mathematical tools to characterize a broad class of scientific uncertainties that arise when available data and credible assumptions are combined to predict population outcomes. As far as this writer is aware, econometric study of partial identification has not been an integral part of philosophical discourse on underdetermination and inductive risk. It warrants attention.

Manski (1995, 2013, 2020, 2024) explain that the motivations for study of partial identification are both principled and practical. On principle, forthright characterization of uncertainty should be a fundamental aspect of the scientific code of conduct. Statistical imprecision and identification problems both limit the conclusions that may be drawn in empirical research. Statistical theory characterizes the inferences that can be drawn about a study population by observing the outcomes of a sample of its members. Studies of identification characterize the inferential difficulties that persist when sample size grows without bound. Identification problems often are the dominant difficulty. Analysis of identification is central to dealing with data quality issues, including missing data and errors in measurement. It is fundamental to understanding the deep problem of distinguishing causation from association.

Forthright characterization of uncertainty serves important societal purposes. Viewing science as a social enterprise, if scientists want people to trust what they say they know, they should honestly state what they don't know. Inferences predicated on weak assumptions can achieve wide consensus, while ones that require strong assumptions may be subject to sharp disagreements.

Despite these motivations, study of partial identification generates strong reactions. The flashpoint of controversy has been the fact that research with weak assumptions generically yields bounds on quantities of interest rather than precise inferences. Some scientists are comfortable reporting findings in the form of bounds and appreciate making explicit the tradeoff between strength of assumptions and strength of findings that the bounds make plain. However, many hold firm to the traditional practice of reporting point estimates and making exact predictions, even though they may rest on fragile foundations. Sections 3 and 4 will discuss two important manifestations of this controversy.

## 1.3. The Lure of Incredible Certitude

The traditional practice is particularly prevalent in economic policy analysis, which attempts to evaluate the impacts of past policies and predict the outcomes of potential future ones. Manski (2011, 2013, 2015, 2020) has criticized policy analysis with *incredible certitude*. Exact predictions of policy outcomes and estimates of the state of the economy are routine. Expressions of uncertainty are rare. Predictions and estimates often are fragile, resting on unsupported assumptions and limited data. Thus, the expressed certitude is not credible.

Manski (2011) introduced a typology of practices contributing to incredible certitude. The typology is

* conventional certitude: A prediction that is generally accepted as true but is not necessarily true.
* dueling certitudes: Contradictory predictions made with alternative assumptions.
* conflating science and advocacy: Specifying assumptions to generate a predetermined conclusion.
* wishful extrapolation: Using untenable assumptions to extrapolate.
* illogical certitude: Drawing an unfounded conclusion based on logical errors.
* media overreach: Premature or exaggerated public reporting of policy analysis.

Some elements of this typology have analogues in philosophy of science. Part of the discussion of "demarcation strategies" in Holman and Whitholt (2022) considers research principles that aim to avoid conflating science and advocacy.

How do researchers motivate reporting findings with incredible certitude? Economists often suggest that researchers respond to incentives. Manski (1995) summarized the status quo this way (p. 3):

> "The scientific community rewards those who produce strong novel findings. The public, impatient for solutions to its pressing concerns, rewards those who offer simple analyses leading to unequivocal policy recommendations. These incentives make it tempting for researchers to maintain assumptions far stronger than they can persuasively defend, in order to draw strong conclusions."

For short, Manski (2020) calls this temptation the *lure of incredible certitude*.

Succumbing to the lure of incredible certitude is not just an affliction of economists. It is common throughout the sciences. Indeed, scientists often seek to justify it by referring to a persistent idea in the philosophy of science, namely that a scientist should deal with underdetermination by choosing one hypothesis among those that are consistent with the available data. To illustrate, I provide two prominent illustrations here.

*1.3.1. Occam's Razor*

To justify selection of one empirically indistinguishable hypothesis out of many, scientists often refer to *Occam's Razor*, the medieval philosophical declaration that "Plurality should not be posited without necessity." Duignan (2017), writing in the *Encyclopaedia Britannica*, gives the usual modern interpretation of this cryptic statement, remarking that: "The principle gives precedence to simplicity; of two competing theories, the simplest explanation of an entity is to be preferred." Swinburne (1997) writes (p. 1):

> "I seek to show that---other things being equal---the simplest hypothesis proposed as an explanation of phenomena is more likely to be the true one than is any other available hypothesis, that its predictions are more likely to be true than those of any other available hypothesis, and that it is an ultimate a priori epistemic principle that simplicity is evidence for truth."

The choice criterion offered here is as imprecise as the one given by Occam. What do Duignan and Swinburne mean by "simplicity?"

Occam's Razor has been viewed favorably not only by many philosophers of science but also by many practicing scientists. The economist Milton Friedman expressed the Occam perspective in an influential methodological essay. Friedman (1953) placed prediction as the central objective of science, writing (p. 5):

> "The ultimate goal of a positive science is the development of a 'theory' or 'hypothesis' that yields valid and meaningful (i.e. not truistic) predictions about phenomena not yet observed."

He later wrote (p. 10):

> "The choice among alternative hypotheses equally consistent with the available evidence must to some extent be arbitrary, though there is general agreement that relevant considerations are suggested by the

> criteria 'simplicity' and 'fruitfulness,' themselves notions that defy completely objective specification."

Thus, Friedman counseled scientists to choose one hypothesis, even though this may require the use of "to some extent . . . arbitrary" criteria. He did not explain why scientists should choose one hypothesis from many. He did not entertain the idea that scientists might offer predictions under the range of plausible hypotheses that are consistent with the available evidence.

Occam's Razor has also played a role in research on artificial intelligence. Pearl (1978) asserted that the (p. 255): "human tendency to regard the simpler as the more trustworthy is given a qualified justification." Blumer et al. (1987) defined an "Occam-Algorithm" and described an (p. 377): "emerging science of machine learning, whose expressed goal is often to discover the simplest hypothesis that is consistent with the sample data." Legg and Hutter (2007) stated that preferring the simplest theory (p. 412): "is generally considered the rational and intelligent thing to do." Hutter et al. (2024) described Occam's razor as (p. 8): "The most important philosophical tool for science."

Some of the artificial-intelligence arguments for simplicity arose from pragmatic concerns related to computational complexity, rather than from the assertion that a simpler model or environment is more likely to be true than a more complex one. Sterkenburg (2025) concluded his argument for simplicity as follows (p. 24): "In this paper, I described a means-ends justificatory argument for Occam's razor from statistical learning theory…It is an honest epistemic justification; and it does not rely on any extra-theoretical assumptions about the true or best hypothesis being simple."

Simplicity may be a well-defined and relevant concept when considering algorithms that vary in computational complexity, but the broader relevance of Occam's razor to science is obscure. The word "simplicity" is vague. In whatever way one may define simplicity, I am not aware of a serious foundation for the belief that simpler environments or hypotheses are more likely to be true than more complex ones.

Dominitz and Manski (2025) observe that (p. 7):

"science gives reason to think that simplicity and complexity co-exist throughout our physical, biological, and social universe. Elementary particles combine to form elements, molecules, planets, and galaxies.

Binary chips connect to form computers and the internet. Organic molecules combine to form cells, plants, and mammals. Individual humans combine to form families, communities, firms, and nations."In each case, understanding the simple building blocks does not suffice to understand the complex aggregate structure. The properties of the structure emerge from the specific spatial manner in which the building blocks combine and connect.

However one tries to operationalize the Occam perspective, its relevance to policy choice and other decisions is not evident. In economics and theapplied science more generally, knowledge of the world and prediction of unobserved phenomena are instrumental to the objective of making good future decisions. Prevailing discussions of Occam's Razor do not pose this objective. Does use of a criterion such as "simplicity" to choose one hypothesis promote good decision making? As far as I am aware, philosophy of science has not addressed this essential question.

#### *1.3.2. Using Statistical Hypothesis Tests to Choose Medical Treatments*

Medical research has endorsed the use of randomized clinical trials to learn the population health outcomes stemming from alternative treatments and to make treatment choices. A common procedure when comparing two treatments in a trial has been to view one as the status quo and the other as an innovation. A statistical hypothesis test is performed, the null hypothesis being that mean response with the innovation and the status quo are the same. If the null hypothesis is not rejected, it is recommended that the status quo treatment continue to be used in clinical practice. If the null is rejected, with the innovation performing better than the status quo in the sample, it is recommended that the innovation become the treatment of choice.

The convention has been to perform a test that fixes the probability of rejecting the null hypothesis when it is correct, the probability of a Type I error. Then sample size determines the probability of rejecting the alternative hypothesis when it is correct, the probability of a Type II error. The power of a test is defined as one minus the probability of a Type II error.

Manski and Tetenov (2016) gave several reasons why hypothesis testing may yield unsatisfactory results for medical decisions and other forms of treatment choice. These are:

*1. Use of Conventional Asymmetric Error Probabilities*: It has been standard to fix the probability of Type I error at 5% and that of Type II error at 10-20%. The theory of hypothesis testing gives no rationale for use of these error probabilities. It gives no reason why a clinician concerned with patient welfare should find it reasonable to make treatment choices that have a substantially greater probability of Type II than Type I error.

*2. Inattention to Magnitudes of Losses to Welfare When Errors Occur*: A clinician should care about more than the probabilities of Type I and II error. A clinician should care as well about the magnitudes of the losses to patient welfare that arise when errors occur. A given error probability should be less acceptable when the welfare difference between treatments is larger. The theory of hypothesis testing does not recognize this.

*3. Limitation to settings with two treatments*: A clinician often chooses among several treatments and many clinical trials compare more than two treatments. Yet the standard theory of hypothesis testing only contemplates choice between two treatments. Statisticians have struggled to extend it to deal sensibly with comparisons of multiple treatments.

In light of these unappealing aspects of hypothesis testing, Manski and Tetenov (2016) argued that a reasonable methodology for treatment choice is the *minimax-regret* criterion, which minimizes the maximum distance from optimality across all possible values of the unknowns in a decision problem. I will explain this criterion in Section 5.

In his discussion of inductive risk, Hempel (1965) essentially recognized the above issues, writing (p. 92): "The problem of formulating adequate rules of acceptance and rejection has no clear meaning unless standards of adequacy have been provided by assigning definite values or disvalues to those different possible 'outcomes' of acceptance or rejection." However, Hempel and other philosophers writing on inductive risk and related ideas have unnecessarily assumed that the only scientifically acceptable alternatives are to accept or reject a hypothesis. For example, Rudner (1953) wrote: "Now I take it that no analysis of what constitutes the method of science would be satisfactory unless it comprised some assertion to the effect that the scientist as scientist accepts or rejects hypotheses."

Neither philosophers of science nor statisticians who study hypothesis testing have recognized that, when the available empirical evidence and other knowledge are consistent with multiple hypotheses, a decision maker need not accept one and reject the others. Combining study of partial identification with reasonable decision criteria, such as minimax regret, yields coherent practical approaches to make medical and other decisions without accepting one among several empirically indistinguishable hypotheses.

### 1.4. Organization of the Paper and Comment on Terminology

To withstand the lure of incredible certitude, partial identification analysis makes set-valued predictions consistent with multiple indistinguishable hypotheses. Section 2 explains basic concepts, placing development of the literature in historical context. Sections 3 and 4 flesh out the concepts by describing two fields of application motivated by important problems of public decision making. These are treatment choice in Section 3 and formation of climate policy in Section 4. Section 5 brings to bear welfare economics and decision theory to recommend reasonable criteria for policy choice with partial identification.

A challenge in writing papers that aim to communicate across disciplines is translation of terminology. Before proceeding, I think it important for readers of this paper to recognize major differences between terminology in philosophy of science and in the disciplines of econometrics and decision theory. It is important that differences in terminology should not impede communication across disciplines that have much to learn from one another.

The philosophical terms *underdetermination* of theories and *inductive risk* have no presence in econometrics and decision theory. Yet they have analogs. Turnbull (2017) differentiates four types of underdetermination. Amongst them, the term *Practical Underdetermination*, which she attributes to Lyre and Eynck (2003), seems particularly close to *partial identification* in econometrics. Turnbull states (p. 5):

> "Practical Underdetermination: In certain theory choices in the practice of science, the evidence which is presently available to us does not direct us to accept one theory over its rivals."

Other related terms include *ambiguity* (Ellsberg, 1961) in decision theory, *Knightian uncertainty* (Knight, 1921) in macroeconomics, *deep uncertainty* in climate science (Flato and Marotzke, 2013), and *model uncertainty* in various disciplines.

Inductive risk is recognizable in statistical hypothesis testing as the occurrence of Type I and Type II errors. Douglas (2000) explicitly relates inductive risk to statistical significance in testing. However, researchers in statistics, econometrics, and decision theory do not use the term inductive risk.

In these fields the word risk, with no adjective attached, has had two interpretations. The most common is mentioned by Hansson (2022) in his *Stanford Encyclopedia of Philosophy* entry on Risk, namely: "the fact that a decision is made under conditions of *known probabilities*." Wald (1950) used the word in a different way in his seminal development of statistical decision theory. Section 5 will argue that the philosophical discussion of inductive risk in Douglas (2000) and other recent work is related to the decision theoretic concept of maximum regret.

## 2. Concepts and History of Partial Identification Analysis

*Econometrics* is the term used by economists to describe study of the logic of empirical inference. Whatever the field of science may be, the logic of empirical inference is summarized by the relationship:

$$\textit{assumptions} + \textit{data} \rightarrow \textit{conclusions}.$$

Econometric analysis has emphasized that data alone do not suffice to draw useful conclusions. Inference requires assumptions (aka theories or hypotheses) that relate available data to a question of interest.

I expect that philosophers of science will immediately associate the above paragraph with the *Duhem-Quine thesis* (Duhem, 1906; Quine, 1951), sometimes called *Holistic Underdetermination* (Turnbull, 2017). The distance between disciplines is well illustrated by the fact that the *Duhem-Quine* nomenclature is famous in philosophy of science but entirely absent from econometrics. In econometrics, the idea that empirical inference requires not only data but also assumptions is considered so self-evident that it is rarely discussed. When it is discussed, it is not attributed to particular scholars.

A fundamental difficulty of empirical research is deciding what assumptions to maintain. I will address this vexing matter in Sections 2.2 and 2.3, after I distinguish two types of scientific uncertainty in Section 2.1. The concepts and history discussed in Section 2 are well-known to econometricians and would not warrant reiteration if they were the audience of this paper. However, I expect that they are largely not familiar to philosophers of science. Hence, I review them here.

### 2.1. Identification and Statistical Imprecision

Researchers often use sample data to learn about theoretically unknown aspects of probabilistic systems. Empirical research confronts problems of statistical imprecision and identification problems. Statistical theory characterizes the inferences that can be drawn by observing a finite sample of outcomes generated by a system. Identification analysis studies inferential difficulties that persist even when sample size grows without bound. The distinction between identification and statistical inference, first formalized by Koopmans (1949), has long been central to econometrics. Koopmans wrote (p. 132):

> "In our discussion we have used the phrase "a parameter that can be determined from a sufficient number of observations. We shall now define this concept more sharply, and give it the name *identifiability* of a parameter. Instead of reasoning, as before, from "a sufficiently large number of observations" we shall base our discussion on a hypothetical knowledge of the probability distribution of the observations, as defined more fully below. It is clear that exact knowledge of this probability distribution cannot be derived from any finite number of observations. Such knowledge is the limit approachable but not attainable by extended observation. By hypothesizing nevertheless the full availability of such knowledge, we obtain a clear separation between problems of statistical inference arising from the variability of finite samples, and problems of identification in which we explore the limits to which inference even from an infinite number of observations is suspect."

Econometricians have subsequently found that separately analyzing statistical inference and identification is highly beneficial to constructive characterization of scientific uncertainty.

A simple example of statistical imprecision occurs when one observes a random sample of population outcomes and uses this average to estimate the population mean outcome. Researchers may measure statistical imprecision of the estimate by its variance or interquartile range, which decrease to zero as sample size increases. Or they may report a confidence set, which shrinks to a point. The Laws of Large Numbers imply that statistical imprecision vanishes as sample size increases to infinity.

Identification problems encompass the spectrum of issues that are sometimes called *non-sampling errors* or *data-quality problems*. These issues cannot be resolved by collecting more data. They may be mitigated only by collecting better data.

A simple yet enormously important example of an identification problem is generated by missing outcome data. Suppose that a random sample of outcomes are drawn, but one does not observe some outcomes. Increasing sample size adds new observations, but it also yields further missing data. If one has incomplete knowledge of the process rendering some outcomes observable and others missing, one cannot precisely learn the population mean outcome as sample size increases to infinity. See Section 2.3.2 for further discussion.

## 2.2. Point and Partial Identification

For most of the twentieth century, econometricians, statisticians, and other researchers studying inference commonly thought of identification as a binary event: a feature of a population is either point-identified or it is not. Empirical researchers combined available data with assumptions that yield point identification, and they reported point estimates, using confidence sets or other devises to measure statistical imprecision. Researchers recognized that point identification often requires strong assumptions that are difficult to motivate, but they saw no other way to perform inference. This perspective parallels the common viewpoint in philosophy of science that underdetermination of theories should be resolved and inductive risk coped with by somehow choosing one hypothesis from the set that are empirically indistinguishable.

Yet there is enormous scope for fruitful inference using weaker and more credible assumptions that partially identify objects of interest. A feature of a population or a parameter in a probabilistic model is *partially identified* if the sampling process and maintained assumptions reveal that the object lies in a set, now called the *identification region* or *identified set*, that is smaller than the logical range of possibilities but larger than a single point.

### 2.2.1. Some History

Prior to about 1990s, analysis of partial identification had a long but scattered history, with contributions by econometricians, other social science methodologists, statisticians, and probability theorists. These early contributions did not use the term partial identification, which came into common use in the early 2000s. Researchers usually referred to their mathematical findings as providing *bounds* on population features or model parameters. I briefly summarize five important early areas of research here.

Perhaps the first well-known contribution was by the econometrician Frisch (1934). He developed sharp bounds on the slope parameter of a simple linear regression when the covariate is measured with mean-zero errors. Fifty years later, his analysis was extended to multiple regression by Klepper and Leamer (1984).

The probability theorist Frechét (1951) studied the conclusions about a joint probability distribution that may be drawn given knowledge of its marginals. Subsequently called *Frechét bounds*, this work formalized an idea suggested one hundred years earlier by the logician Boole (1854). See Ruschendorf (1981) for subsequent findings.

The sociologists Duncan and Davis (1953) used a numerical example to show that *ecological inference*, which concerns decomposition of a probability mixture into its components, is a problem of partial identification. Formal characterization of identification regions did not occur until close to half a century later, in Horowitz and Manski (1995) and Cross and Manski (2002).

The statisticians Cochran, Mosteller, and Tukey (1954) suggested conservative finite sample analysis of surveys with missing outcome data due to nonresponse by sample members. They did not analyze identification directly. Instead, they considered the maximum mean square error of an estimate of a population mean when nothing is known about the process generating missing data. Analysis of maximum bias is related to analysis of partial identification but not identical, as it combines study of identification and statistical imprecision. Cochran (1977) subsequently downplayed their early suggestion, opining that the values obtained for maximum mean square error were too large to be useful. Formal partial identification analysis began in Manski (1989).

The epidemiologist Peterson (1976) initiated study of partial identification of the competing risk model of survival analysis. Crowder (1991) and Bedford and Meilijson (1997) carried this work further.

2.2.2. Development of the Modern Econometric Literature

Despite the above and other early contributions, the concept of partial identification long remained at the fringes of scientific consciousness and did not spawn systematic study. A coherent body of econometric research began to take shape in the 1990s, developing to the degree that study of partial identification is now a mainstream subfield of econometrics. A sequence of in-depth reviews of the literature, written at different technical levels and with different emphases, include Manski (2003, 2007), Tamer (2010), Canay and Shaikh (2017), and Molinari (2020). Statisticians, epidemiologists, and computer scientists have also made contributions. However, analysis of partial identification continues to be at the fringe rather than the mainstream of those disciplines.

Part of the modern literature on partial identification in econometrics focuses on the identification problems generated by imperfect data quality, including measurement error and missing data. Part focuses on identification of structural econometric models used to describe human behavior and interactions. Another part focuses on analysis of treatment response, a subject that some social scientists refer to as "causal inference."

Whatever the specific subject under study, a common theme runs through the literature. One first asks what available data combined with relatively weak and highly credible assumptions reveal about the population or probabilistic phenomenon of interest. One then studies the identifying power of stronger assumptions that still aim to retain acceptable credibility.

### 2.2.3. Interpreting Credibility

I have thus far used the words *credible* and *credibility* without definition. There appears to be consensus in philosophy of science and across the sciences that c*redibility* is a fundamental subjective concept connoting belief regarding the truth of an assumption or theory. Yet the concept may defy deep definition.

When discussing incredible certitude in Section 1, I wrote that predictions and estimates resting on unsupported assumptions and limited data are not credible. Symmetrically, I think it appropriate to now say that predictions and estimates are credible if they rest on well-supported assumptions and sufficient data. However, this interpretation of credibility is superficial. What are meant by "well-supported assumptions" and "sufficient data?"

Manski (2013, page 12) observed that the Second Edition of the *Oxford English Dictionary* (OED) defines *credibility* as "the quality of being credible," defines *credible* as "capable of being believed; believable," and defines *believable* as "able to be believed; credible." Thus, the OED comes full circle, unable to explain what credibility means.

Rather than attempt a general definition of credibility, philosophers of science often develop and discuss evocative examples. A well-known case is the *grue-bleen* illustration used by Goodman (1955) to exemplify what he called a *new riddle of induction*. Existing data alone cannot predict whether objects that been observed in the past to have blue and green colors respectively will continue to have these colors in the future or will switch to green and blue. Nevertheless, a scientist who knows color to be a manifestation of the electromagnetic spectrum and who assumes the physical properties of objects to be stable over time might reasonably find the first prediction credible and the second not credible.

Econometric research on partial identification has emphasized that credibility is more than a binary concept. Viewing credibility in an ordinal way, Manski (2003) introduced a principle called (p. 1):

> "*The Law of Decreasing Credibility*: The credibility of inference decreases with the strength of the assumptions maintained."

This principle does not prescribe how scientists should assess degree of credibility. It simply cautions researchers that they face a logical dilemma as they decide what assumptions to maintain: Stronger assumptions yield inferences that may be more powerful but less credible.

An important objective of modern partial identification analysis is to determine the conclusions one can draw in empirical research without imposing untenable assumptions. Manski (1995) argued that performing inference with weak assumptions can establish a domain of consensus among researchers who may hold disparate beliefs about what assumptions are appropriate. It makes plain the limits to credible inference.

Writers on partial identification recommend that when identification regions obtained with available data and well-supported assumptions turn out to be large, researchers should face up to the fact that they cannot draw conclusions as strong as they might like to achieve. This recommendation differs fundamentally from that of Occam's Razor, statistical hypothesis testing, and other devices commonly used to choose one of multiple empirically indistinguishable hypotheses.

## 2.3. A Continuing Controversy: Coping with Missing Data

Notwithstanding the development of a substantial modern body of research on partial identification, drawing set-valued conclusions based on well-supported assumptions is controversial among economists, as well as in other disciplines. The lure of incredible certitude remains strong. Many researchers continue to think that empirical research is useful only if it studies point identified quantities.

### 2.3.1. Assuming that Data are Missing at Random

A continuing controversy of considerable practical importance, which should be of interest to philosophy of science, concerns dealing with missing data in empirical research. A prominent manifestation of incredible certitude is the widespread use of the assumption that missing outcome or covariate data are *missing at random (MAR)*. Formally, this assumption asserts that the process generating missing outcome or covariate data is statistically independent of actual outcome values. When data are assumed to be MAR, analysis of observed data commonly yields point identification of population features and model parameters. Social scientists often use the MAR assumption to impute missing data values, as recommended by Rubin (1987) and Little and Rubin (2019).

MAR assumptions may appear simple, but they rarely are credible when studying *observational data*; that is, data generated by passive observation of a population. They moreover are rarely credible when studying data generated by imperfect randomized experiments, in which some subjects do not comply with assigned treatments or are lost to follow up before their outcomes are observed.

Perhaps the singular circumstance in empirical research with scientific consensus that the MAR assumption is highly credible is when studying data from an ideal randomized experiment, with complete compliance and follow up. Missing data are inevitable in experiments because subjects assigned to one treatment group only experience the outcome of that treatment. The outcomes that subjects would receive with other treatments are counterfactual, hence necessarily unobservable. In an ideal experiment, randomized assignment to treatment makes counterfactual outcomes MAR.

### 2.3.2. Agnostic Partial Identification Analysis

The most basic partial-identification prescription for empirical research with missing data is to be completely agnostic about the process generating missingness. Agnostic analysis was proposed by Manski (1989) in a mathematically simple setting where the objective is to learn the population distribution function or the mean of a real-valued outcome in a population. It has subsequently been generalized to more complex settings. The objective may be to learn a quantile, a spread parameter such as the variance or interquartile range, or the parameters of a specified model. In regression analysis, data on covariates as well as outcomes may be missing. Manski (2003, 2007), Tamer (2010), and Molinari (2020) review the literature.

To appreciate the basic idea of agnostic analysis, it suffices to consider the problem of learning the distribution function of an outcome. Suppose that data are generated by random sampling, but some outcomes are not observed. The implications are seen immediately by use of the Law of Total Probability.

Let y denote the outcome and P(y) be its population distribution. Let z be a binary indicator, with $z = 1$ if an outcome is observable and $z = 0$ otherwise. For any real number d, the probability that y is less than or equal to d is

$$P(y \leq d) \;=\; P(y \leq d|z = 1)\cdot P(z = 1) + P(y \leq d|z = 0)\cdot P(z = 0).$$

As sample size goes to infinity, random sampling reveals $P(y \leq d|z = 1)$, the probability that observable values of y are less than or equal to d. It also reveal $P(z = 1)$ and $P(z = 0)$, the probabilities that an outcome is observable or missing. Data collection reveals nothing about $P(y \leq d|z = 0)$, the probability that missing outcomes are less than or equal to d. In the absence of assumptions, this probability is only known to lie in the interval [0, 1]. Hence, agnostic analysis yields this informative sharp bound on $P(y \leq d)$:

$$P(y \leq d|z = 1)\cdot P(z = 1) \;\leq\; P(y \leq d) \;\leq\; P(y \leq d|z = 1)\cdot P(z = 1) + P(z = 0).$$

The bound has width $P(z = 0)$, the probability with which outcomes are missing.

Being agnostic about missing data provides a logical starting point for empirical research, yielding inferences with maximal credibility. Nevertheless, researchers typically would like to draw stronger conclusions. The MAR assumption yields point identification by asserting that $P(y \leq d|z = 0) = P(y \leq d|z = 1)$, but the cost in credibility may be high. Research on partial identification explores the vast middle ground between making no assumptions and ones that yield point identification. Section 3.2 will discuss an important class of middle-ground assumptions.

## 3. Identification of Treatment Response

I now explore further the continuing controversy regarding the conduct of empirical research with missing data. Missing data are inevitable in analysis of treatment response, regardless of whether treatments are assigned randomly in a trial or through some other process. Outcomes are potentially observable only for treatments that persons actually receive. The outcomes that persons would receive with other treatments are counterfactual, and hence must be unobservable.

Social scientists who consider point estimation to be essential but who lack data from ideal randomized experiments often perform *design-based inference* of observational data or imperfect experiments. Such inference produces point estimates of concepts of treatment effects that may be credibly point-identified but that commonly lack substantive relevance. In contrast, partial identification analysis yields set-valued estimates of concepts of treatment effects that aim to inform future treatment choice using the welfare-economic paradigm. Sections 3.1 and 3.2 explain.

### 3.1. Design-Based Inference

Modern advocacy of design-based inference began with the work of Donald Campbell and his collaborators; e.g., Campbell and Stanley (1963). Campbell distinguished between the internal and external validity of studies of treatment response. A study is said to have *internal validity* if it yields credible findings for the study population. It has *external validity* if the findings may be credibly extrapolated to a population of substantive interest.

Campbell argued that studies should be judged primarily by their internal validity and secondarily by their external validity. This perspective has been used to argue for the primacy of experimental research over observational studies, whatever the study population may be. The appeal of an ideal randomized experiment is its internal validity. Such experiments have no inherent advantage in terms of external validity.

Campbell's doctrine of the primacy of internal validity has been extended from randomized experiments to observational studies. When considering the design and analysis of observational studies of treatment response, Campbell and collaborators recommended that researchers aim to emulate as closely as possible the conditions of an ideal randomized experiment, even if this requires focus on a study population that differs substantially from the population of interest. This has led to development of methodologies for analysis of so-called *quasi-experiments*. These approaches to study of observational data have been promoted for their internal validity, leaving external validity as a secondary concern.

Thistlethwaite and Campbell (1960) introduced *regression-discontinuity* analysis of policies that use observable institutional rules to determine treatment assignment. Such analysis may credibly point-identify treatment response in the sub-population of persons who are close to the threshold determining treatment assignment. They do not yield credible conclusions for persons distant from the threshold.

Adapting econometric research on linear fixed-effects models of panel data (e.g., Mundlak, 1978), applied microeconomists have increasingly performed *difference-in-difference* analyses of temporal changes in policy (e.g., Card and Krueger, 1994). Such analysis maintains a questionable *parallel trends* assumption. Even when this assumption is credible, it only identifies average treatment effects in treatment units where treatment assignment changed over time.

Since the mid-1990s, the Campbell perspective has been championed by applied microeconomists who advocate study of a certain *local average treatment effect* (LATE). This is defined as the average treatment effect within the sub-population of so-called *compliers*, these being persons whose received treatments would be modified by hypothetically altering the value of an instrumental variable; see Imbens and Angrist (1994). Local average treatment effects are not policy-relevant quantities; see the discussions in Manski (1996, 2007), Deaton (2010), and Heckman and Urzua (2010). Their study has been motivated by the fact that they are point-identified given particular assumptions that are sometimes thought credible. Angrist and Pischke (2010) call research using the LATE concept and other types of design-based inference a *credibility revolution.*

Research on LATE exemplifies the widespread reluctance of researchers to face up to uncertainty in policy analysis. Researchers often are aware that they cannot form a credible point estimate of a policy-relevant treatment effect. They could face up to uncertainty and determine what they can credibly infer about this quantity, perhaps obtaining an informative bound. Instead, they change the objective and focus on a different concept of treatment effect that is not of substantive interest but that can be point-estimated credibly. Thus, they sacrifice relevance for certitude.

Notable scientists have critiqued sacrificing relevance for certitude, but it persists. John Tukey wrote (Tukey, 1962, p. 13-14): “Far better an approximate answer to the right question, which is often vague, than an exact answer to the wrong question, which can always be made precise.” Many cite versions of the joke about the drunk and the lamppost. Noam Chomsky has been quoted as putting it this way (Barsky, 1998, p. 95): “Science is a bit like the joke about the drunk who is looking under a lamppost for a key that he has lost on the other side of the street, because that's where the light is." Kaplan (1964) stated it this way (p. 11): "There is a story of a drunkard searching under a street lamp for his house key, which he had dropped some distance away. Asked why he didn't look where he had dropped it, he replied, 'It's lighter here!' "

Why are many researchers content to look under a lamppost? It is rare to find an explicit rationale. Scientists sometimes motivate research as an effort to improve our "understanding" of a subject or to perform credible "causal inference." They argue that these are worthwhile objectives even if there are no practical implications for decision making. From the perspective of philosophy of science, design-based inference proceeds in the tradition of value-free science, independent of society.

### 3.2. Partial Identification of Mean Treatment Response

Partial identification analysis of treatment response begins by posing an objective of substantive concern, whose accomplishment requires quantitative knowledge of some concept of treatment response. One determines the conclusions about this concept achievable by combining available data and credible assumptions. The generic finding is a set of possible values.

An objective of considerable practical concern considers a social planner who uses the welfare-economic paradigm to choose a treatment for each member of a population. A central case is medical decision making, in which a health planner or clinician chooses treatments for a population of patients. Let J list the members of the population. Let T denote a set of alternative feasible treatments. For each $j \in J$ and $t \in T$, let $y_j(t)$ be a real outcome that person j would experience with treatment t. For example, t = A may be surveillance and t = B an aggressive treatment. Outcome $y_j(t)$ may express patient health resulting from both illness and side effects.

It is common to study a simple form of the planner's choice problem in which these conditions hold:

(1) Treatment is *individualistic*, meaning that the treatment received by one person does not affect the outcomes experienced by others.

(2) Every member of J must receive the same treatment.

(3) The social welfare function aims to choose a treatment that maximizes the mean outcome.

Then empirical research should seek to learn the mean outcomes $E[y(t)]$, $t \in T$.

Medical research studies more complex versions of this setting, particularly ones in which patients may differ in their observable attributes and the health planner or clinician can personalize treatment, making choices vary with patient attributes. The discussion below generalizes to these settings.

3.2.2. Agnostic Analysis

Analysis of treatment response has commonly supposed that a researcher observes the treatments received and outcomes realized by a random sample of the population of interest. Let z be the treatment received by a member of the population and let $y = y(z)$ be the realized outcome. Let $P(z = t)$ be the fraction of persons who receive t.

Manski (1990) performed agnostic partial identification analysis of mean treatment response. The starting point is the Law of Iterated Expectations, which shows that for each treatment t,

$$E[y(t)] \;=\; E[y(t)|\, z = t]\cdot P(z = t) + E[y(t)|z \neq t]\cdot P(z \neq t).$$

Data collection can reveal $P(z = t)$ and the mean observable outcomes $E[y(t)|z = t]$. It cannot reveal the mean counterfactual outcome $E[y(t)|z \neq t]$. If treatment outcomes may be unbounded, $E[y(t)|z \neq t]$ can take any value between minus and plus infinity. Hence, agnostic analysis is uninformative about $E[y(t)]$.

The situation differs qualitatively if outcomes are bounded, as they often are in empirical research. For concreteness, suppose that $y(t)$ must lie in the interval $[0, 1]$. Then $E[y(t)|z \neq t]$ must lie in this interval. Hence, agnostic analysis yields this informative sharp bound on $E[y(t)]$:

$$E[y(t)|z = t]\cdot P(z = t) \;\leq\; E[y(t)] \;\leq\; E[y(t)|z = t]\cdot P(z = t) + P(z \neq t).$$

The bound has width $P(z \neq t)$, the probability with which outcomes are counterfactual. This agnostic finding is analogous to the one discussed in Section 2.3.1.

3.2.2. Analysis with Instrumental Variables

Agnostic analysis provides a logical starting point for research on treatment response, having maximal credibility. An MAR assumption yields point identification by asserting that $E[y(t)|\ z \neq t) = E[y(t)|z = t]$. but the cost in credibility may be high. Beginning with independent work by Manski (1990) and Robins (1989), a prominent subject of study in research on partial identification has been the identifying power of assumptions that use an *instrumental variable* (IV). Swanson *et al*. (2018) reviews part of the literature.

An IV is an observable covariate, say v. Reiersol (1945) thought of such a covariate as an instrument to help to identify an object of interest. Early econometricians used IVs to point-identify linear structural equation systems. Modern econometric research uses IVs to help address many identification problems, including identification of treatment response. Observation of an IV is not useful per se. It is useful only when combined with an assumption that has identifying power.

It is particularly common to assume that different sub-populations of a study population experienced different processes of treatment selection but share the same distribution of treatment response, or at least the same mean response. For example, when studying patient care, one might assume that patient sub-populations treated by different clinicians have the same distribution of treatment response. Then the identity of the clinician is the IV.

An assumption that different sub-populations share the same distribution of treatment response, or same mean response, generates an *intersection bound* on treatment response. Consideration of each sub-population separately yields an agnostic bound using no knowledge of treatment allocation. The assumption of a common distribution of treatment response across sub-populations implies that the distribution must simultaneously lie in each agnostic bound. Thus, it must lie in the intersection of the bounds.

For example, suppose that one is studying patient care, and that data are available for three clinicians. Let the outcome of interest be patient survival, which may take the value 1 or 0. Suppose that, considering each clinician separately, one obtains these clinician-specific bounds on the probability of survival: [0.4, 0.7], [0.2, 0.6], [0.5, 1]. Suppose that one finds it credible to assume survival probabilities are the same for the patients treated by each clinician. Set intersection yields [0.5, 0.6] as the bound on survival probability obtained by combining data across the three clinicians.

The example demonstrates a subtlety in the operation of set intersection. The result of set intersection depends on the joint positioning of sets, not only their sizes. One might think that when multiple studies yield identification regions of different sizes, the studies yielding the smallest identification regions would play the most prominent role when combining findings. The example shows that this need not be the case. The lengths of the three clinician-specific intervals are 0.3, 0.4, and 0.5 respectively. Yet the set intersection is determined entirely by the second and third intervals.

Intersection bounds implied by assumptions using IVs exemplify the Law of Decreasing Credibility. Assuming that different sub-populations share the same distribution of treatment response often has identifying power, narrowing the identification region obtained with agnostic analysis. Strengthening the conclusions drawn in empirical research is useful if the maintained IV assumption is sufficiently credible, but it may be misleading if the assumption lacks credibility.

## 4. Partial Identification of Climate Models

One might think that econometrics and climate modeling are so different that econometric methodology would not be relevant to climate modeling. However, partial identification analysis can be constructively applied to climate models, serving to frame and interpret their uncertainties. This section draws on discussion in Manski, Sanstad, and DeCanio (2021).

First, research in both econometrics and climate modeling faces broadly similar identification problems due to data imperfections. Climate data that might have been available in principle may have measurement errors or be missing in practice, especially data documenting climate dynamics prior to the existence of modern instrumentation. Moreover, climate data in counterfactual settings is unavailable in principle. We cannot observe what climate change would have occurred over the past century if the trajectory of greenhouse gas emissions had differed from its actual path.

Second, research in both domains faces the problem of modeling a complex system. The climate system comprises many different physical processes and mechanisms occurring at a range of spatial and temporal scales. Designing and implementing a climate model requires many choices regarding the underlying architecture, methods of numerical approximation, solution technique, which physical processes to include as model components, and how to represent these processes. Climate model uncertainty arises from incomplete scientific understanding of physical phenomena, limitations of quality and quantity of empirical data available to fit models, and the need to make approximations for computational tractability.

The highly complex nature of the climate system makes it inevitable that the computational models used to represent it are subject to a degree of irreducible imprecision. Although the basic physics of climate science is common across models, the way that the physics is expressed in a particular model is subject to numerous practical choices of implementation, including choice of coordinate system, discretization approach, and numerical solutions. In addition, climate models differ in the details of which physical processes are selected for inclusion and how they are represented. The multiplicity of models and inter-model variations of the quantitative outputs they generate are referred to by climate scientists as reflecting "structural uncertainty".

### 4.1. Multi-Model Ensemble Analysis

To quantify and analyze structural uncertainty, sets of models are used to study how different structural choices affect climate predictions. This is called *multi-model ensemble* (MME) analysis.

MMEs are based on model intercomparison projects in which the characteristics and performance of multiple models are analyzed. These studies examine the individual and collective ability of models to simulate the current and historical climate, analyze the factors driving model behavior and inter-model differences, and provide a means of assessing changes in model performance over time.

Despite considerable effort, there continue to be shortcomings in the models' accuracy and precision. Moreover, it has not proven possible to unambiguously rank models with respect to performance or to identify a "best" model using historical climate information. The extent of structural uncertainty can be gauged by the fact that there are currently several dozen different modeling groups around the world, each running one or more versions of its own model.

Abramowitz et al. (2018) take the view that different models can be thought of as representing different "working hypotheses" about how to best represent the details of the structure and dynamics of the climate. Parker (2006) characterizes the climate modeling community's acceptance of multiple, co-existing models as "model pluralism," with different models seen as complementary. Thus, climate models are only partially identified.

Model pluralism has generally been maintained in MME studies of the future climate, in that the sets of models used in specific studies have usually been determined simply by which modeling teams choose to participate, a procedure that Knutti (2010) describes as "model democracy." That available models are not assessed or screened for inclusion essentially reflects the unavailability of reliable, agreed-upon criteria for doing so: If structural uncertainty (partial identification) cannot be resolved through standard evaluation and performance intercomparison, there are no obvious grounds to omit particular models from climate projection studies.

A number of methods have been developed and applied to MME uncertainty quantification and analysis. Virtually all methods combining MME outputs into single projected climate trajectories. There appear to be two primary reasons that this form of model combination predominates. One is that it has been found that averaging point estimates of observed past climate change across models can improve accuracy relative to the average accuracy of the point estimates produced by individual model. The other is that modelers perceive policy-makers as requiring single projections (as functions of individual GHG emissions scenarios) for use in decision-making.

As this type of work has proceeded, climate scientists and others have recognized persistent methodological problems in combining model projections. Regarding the finding that accuracy averaging model estimates can improve upon the average accuracy of estimates of individual models, it has been shown that this may be the consequence of an algebraic relationship rather than an empirical result.[2] Moreover, model performance in historical data has not been demonstrated to imply skill in predicting the future climate.

Combining climate model ensemble outputs into single projected trajectories of the future global climate remains a challenging and unresolved problem. Observing this situation, Manski, Sanstad, and DeCanio (2021) show that viewing model structural uncertainty in terms of partial identification facilitates an alternative approach to using MME information in climate policy formation. Recognizing that there currently exists no "best" model, they argue that climate scientists should use multiple credible models to forecast future global climate. Choice of climate policy would recognize that each such model may be correct. Manski, Sanstad, and DeCanio (2021) applied the minimax-regret criterion to operationalize this idea, using an integrated assessment model to quantify the policy tradeoff between abatement cost and the societal damage from climate change.

## 5. Welfare-Economic Policy Choice with Partial Identification

As far as I am aware, philosophy of science has largely studied underdetermination of theories and inductive risk as separate subjects. Prior to about 2000, underdetermination was commonly addressed from the perspective of value-free science: a concern of science per se rather than of the society that uses science to inform decisions. In Section 1, I noted that advocacy of Occam's Razor in the philosophy and practice of science has not sought to explain the relevance of the Occam principle to decision making. Underdetermination has been connected to values since 2000, as in Trumbull (2017), but generally without formal study of the use of science in societal decision making.

[2] Jensen's Inequality implies that when mean square error or another convex loss function is used to measure predictive accuracy, the accuracy of a simple or weighted mean prediction is algebraically better than the corresponding average accuracy of individual model predictions. See Manski (2016) for detailed discussion.

In contrast, discourse on inductive risk has emphasized how scientific uncertainty affects decisions. For example, the Douglas (2000) discussion of errors in statistical hypothesis testing states (p. 567):

> "In setting the standard for statistical significance, one must decide what balance between false positives and false negatives is optimal. In making this decision, one ought to consider the consequences of the false positives and false negatives, both epistemic and non-epistemic. . .
>
> In testing whether dioxins have a particular effect or not, an excess of false positives in such studies will mean that dioxins will appear to cause more harm to the animals than they actually do, leading to overregulation of the chemicals. An excess of false negatives will have the opposite result, causing dioxins to appear less harmful than they actually are, leading to underregulation of the chemicals."

Douglas and other philosophers writing on inductive risk recognize that methodological choices made in statistical hypothesis testing can affect public policy.

Research in econometrics and decision theory proposes and analyzes criteria for reasonable decision making with partial identification, thus unifying study of underdetermination and inductive risk. An early prominent idea, with origins in the 1700s, is Bayesian decision theory. Applied to welfare-economic analysis of public policy, a Bayesian planner places a subjective distribution on all unknown quantities and makes a decision that maximizes subjective expected social welfare.

Modern research on decision making with partial identification largely studies decision making under ambiguity; that is, in settings where the decision maker does not find it credible to place a subjective probability distribution on unknowns. The connection between partial identification and decision under ambiguity was made in Manski (2000), writing (page 416):

> "This paper connects decisions under ambiguity with identification problems in econometrics. Considered abstractly, it is natural to make this connection. Ambiguity occurs when lack of knowledge of an objective probability distribution prevents a decision maker from solving an optimization problem. Empirical research seeks to draw conclusions about objective probability distributions by combining assumptions with observations. An identification problem occurs when a specified set of assumptions combined with unlimited observations drawn by a specified sampling process does not reveal a distribution of interest. Thus, identification problems generate ambiguity in decision making."

This use of the term *ambiguity* follows Ellsberg (1961), who used the word to signify uncertainty when one specifies a set of feasible states of nature but does not place a probability distribution on the state space.

To conclude this paper, I explain basic concepts and specific criteria that can be applied to societal decision problems such as those described in Sections 3 and 4, using the welfare-economic paradigm. These concepts and criteria do not require a social planner making a policy choice with scientific uncertainty to accept one hypothesis and reject all others. To the contrary, they enable a planner confronting underdetermination to coherently consider a set of possible truths. The present discussion draws on Manski (2024), Chapter 1.

### 5.1. Basic Concepts of Decision Theory

Decision theory studies principles for reasonable decision making under uncertainty. To begin, one supposes that outcomes are determined by the chosen action and by some feature of the environment, called the *state of nature*. The decision maker is assumed able to list all states of nature that could possibly occur. This list, called the *state space*, provides the basic expression of uncertainty.

The state space is a list of 'known unknowns.' Decision theory supposes that the decision maker does not contemplate the possible existence of unlisted 'unknown unknowns.' Thus, decision theory does not attempt to grapple with the hugely difficult possibility of *unconceived alternatives*, discussed in the philosophy of science literature by Stanford (2006).

A state of nature signifies the possible truth of a scientific theory or other property of the decision environment. The state space contains only one element when a decision maker uses Occam's Razor or some other device to accept one theory out of many. It contains multiple elements when the decision maker lacks a justifiable basis to reject all but one theory. In the philosophical language of underdetermination, the state space lists the set of credible theories. In econometric language, it is the identification region. The larger the state space, the greater the scientific uncertainty.

A fundamental difficulty in decision making under uncertainty is apparent even in a simple setting with two feasible actions, say A and B, and two possible states of nature, say $s_1$ and $s_2$. Suppose that a planner or other decision maker wants to maximize a specified welfare function. Action A yields higher welfare in state $s_1$ and B higher welfare in $s_2$. If it is not known whether $s_1$ or $s_2$ is the true state, it is not known which action is better. Thus, maximization of welfare is logically impossible. At most one can seek a reasonable way to make a choice. A basic issue is how to interpret and justify the word "reasonable."

Two conceptually distinct but mathematically related approaches have been used to develop criteria for reasonable decision making. *Consequentialist* decision theory focuses on the substantive consequences of choices. It offers prescriptions for reasonable future decision making under uncertainty. *Axiomatic* theory poses choice axioms deemed intellectually appealing as primitives and proves theorems showing that behavior consistent with the axioms is equivalent to using particular classes of consequentialist decision criteria. Thus, the theorems seek to interpret choices, not to predict or prescribe them. I focus here on consequentialist theory.[3]

---

[3] The representation theorems of axiomatic theory are if-and-only-if statements, proving that adherence to certain choice axioms implies particular classes of consequentialist decision criteria, and vice versa. For example, the famous axioms of Savage (1954) hold if and only if a decision maker maximizes subjective expected utility. The mathematical connection between axiomatic and consequentialist theory may suggest that the two perspectives are equivalent, but this is not so. Consider the interpretation of welfare and uncertainty in consequentialist expected utility theory and in the Savage axiomatic theory. In the former, the welfare function and the subjective probability distribution are primitives specified by the decision maker, expressing what the decision maker wants to achieve and their perception of uncertainty. In the latter, the welfare function and subjective distribution are mathematical constructs implied by a pattern of hypothetical choice behavior. From the consequentialist perspective, a welfare function and subjective

5.2. Consequentialist Decision Theory

Consequentialist decision theory specifies a welfare function, a choice set, and a state space as primitives. It then seeks reasonable criteria to make decisions. A recommendation made as early as the late 1700s has been maximization of expected utility. One places a subjective probability distribution on the state space and chooses an action that maximizes the expected value of welfare with respect to this distribution.

To assist decision makers who do not find it credible to express uncertainty through a probability distribution, modern decision theorists study criteria that, in some sense, works uniformly well over all of the state space. Two prominent interpretations of this idea are the maximin and minimax-regret criteria. I formalize these concepts presently.

5.2.1. The Choice Set, State Space, and Welfare Function

In mathematical terms, decision theory supposes that a social planner or other decision maker faces a predetermined choice set C and assumes that the true state of nature $s^*$ lies in a state space S. The welfare function $w(\cdot, \cdot): C \times S \rightarrow R^1$ maps actions and states into welfare. The welfare function expresses the values of the decision maker. The decision maker wants to maximize $w(\cdot, s^*)$ over C, but does not know $s^*$.

The state space is subjective, but it is not an arbitrary construction. Credibility is a fundamental matter in consequential decision theory. If decisions aim to enhance welfare in the real world, the decision maker should specify a state space that embodies some reasonable sense of credibility.

Research seeks to provide at least a partially objective basis for specification of the state space. This basis is obtained by combining plausible theory with available data. This chain of thought directly connects

---

distribution aim to express a decision maker's goals and uncertainty in a realistic manner. In the Savage theory, they have no necessary relationship to an objective reality.

identification analysis to consequentialist decision theory. The identification region derived in econometric study of partial identification aims to provide a credible state space for decision making.

The choice set, state space, and welfare function constitute the basic formal concepts of consequentalist decision theory. An additional concept comes into play when the decision maker observes finite-sample data drawn from some sampling process. Then the sampling process and the extent of data collection matter. Wald (1950) extended decision theory to settings with sample data. This extension requires only a modest generalization of concepts, but it is mathematically more complex. I first discuss decisions without sample data.

### 5.2.2. Decisions without Sample Data

There is consensus among decision theorists that decisions should respect dominance. Action $c \in C$ is weakly dominated if there exists a $d \in C$ such that $w(d, s) \geq w(c, s)$ for all $s \in S$ and $w(d, s) > w(c, s)$ for some $s \in S$. The fundamentally hard part of decision making under uncertainty is choice among undominated actions.

To choose among undominated actions, decision theorists have proposed various ways of using $w(\cdot, \cdot)$ to form functions of actions alone, which can be optimized. Three criteria have been particularly prominent. They are

*Maximization of Subjective Expected Welfare:* Suppose that one finds it credible to place a subjective probability distribution, say $\pi$, on the state space. Then one averages state-dependent welfare with respect to $\pi$, and maximizes subjective average welfare over C: Thus, one solves the problem

$$\max_{c \in C} \int w(c, s) d\pi.$$

*Maximin*: In the absence of a subjective distribution on S, a prominent idea is to choose an action that, in some sense, works uniformly well over all of S. One interpretation of this idea is to maximize the minimum welfare attainable across S. The maximin criterion solves the problem

$$\max_{c \in C} \ \min_{s \in S} \ w(c, s).$$

*Minimax Regret*: A different interpretation of making a choice that works uniformly well is the minimax-regret criterion, which solves the problem

$$\min_{c \in C} \ \max_{s \in S} \ [\max_{d \in C} w(d, s) - w(c, s)].$$

The expression $\max_{d \in C} w(d, s) - w(c, s)$ is called the *regret* of action c in state s. Regret measures the degree of sub-optimality of action c if s is the true state of nature. One does not know the true state, so the prescription is to evaluate c by its maximum regret over S and select an action that minimizes maximum regret.

The maximin and minimax regret criteria are sometimes confused with one another, but they yield the same choice only in certain special cases. The maximin criterion considers only the worst outcome that an action may yield. Minimax regret considers the worst outcome relative to the best achievable in a given state of nature. It quantifies how lack of knowledge of the true state of nature diminishes the quality of decisions.

*Illustration*

To illustrate how maximin and minimax-regret choices may differ, consider a planner's welfare-economic choice between two treatments A and B as discussed in Section 3. Suppose that, using credible assumptions, the identification region for mean treatment response under treatments A and B is the bounded rectangle $[L_A, U_A] \times [L_B, U_B]$, where L and U denote lower and upper bounds. Suppose that neither treatment dominates; that is, $U_B > L_A$ and $U_A > L_B$. Thus, it is uncertain which treatment is better.

If the planner must assign everyone to the same treatment, the maximin criterion selects treatment A if $L_A > L_B$ and B if $L_A < L_B$. The minimax-regret criterion selects A if $U_B - L_A < U_A - L_B$ and B if $U_B - L_A > U_A - L_B$.

Manski (2009) studies a more subtle setting where the planner has the power to diversify treatment, assigning some fraction of the population to one treatment and the remaining fraction to the other. Then the maximin treatment choice remains as above, but minimax-regret diversifies treatment. It assigns the fraction $(U_B - L_A)/[(U_B - L_A) + (U_A - L_B)]$ to treatment B and the remainder to A.

The broad argument for diversification is that it enables a planner to balance two types of potential error. A Type A error occurs when treatment A is chosen but is actually inferior to B, and a Type B error occurs when B is chosen but is inferior to A. Assigning everyone to treatment A entirely avoids type B errors but may yield Type A errors, and vice versa for assigning everyone to treatment B. Fractional allocations make both types of errors but reduce their potential magnitudes. In colloquial terms, they avoid the danger of "putting all eggs in one basket."

### 5.2.3. Decisions with Sample Data

Analysis of statistical decision problems add to the above structure by supposing that the decision maker observes data generated by some sampling distribution. Knowledge of the sampling distribution is generally incomplete. To express this, one extends state space S to list the feasible sampling distributions.

Wald (1950) defined a statistical decision function (SDF) to be any function $c(\psi)$ that maps observed sample data $\psi$ into an action. His concept of a statistical decision function embraces all mappings [data $\rightarrow$ action]. An SDF is a deterministic function after realization of the sample data, but it is a random function ex ante. Hence, the welfare achieved by an SDF is a random variable ex ante. Wald's theory evaluates the performance of an SDF in state s by the ex-ante distribution of loss that it yields across realizations of the sampling process. Thus, statistical decision theory resides with the standard framework of frequentist statistical theory.

Wald proposed measurement of the performance of an SDF in state s by the mean welfare it yields across samples. (Using a mathematical setup where one wants to minimize loss rather than maximize welfare, he referred to mean loss across samples as *risk*.) This done, it is simple to generalize the three decision criteria discussed above to encompass decision making with sample data. One replaces each occurrence of the welfare $w(c, s)$ with mean welfare $E\{w[c(\psi), s]$ across repeated samples and then applies a criterion. Thus, the statistical versions of the above criteria are

$$\max_{c(\cdot) \in \Gamma} \int E_s\{w[c(\psi), s]\}\, d\pi,$$

$$\max_{c(\cdot) \in \Gamma} \min_{s \in S} E_s\{w[c(\psi), s]\},$$

$$\min_{c(\cdot) \in \Gamma} \max_{s \in S} \left( \max_{d \in C} w(d, s) - E_s\{w[c(\psi), s]\} \right).$$

where $\Gamma$ denotes a specified set of feasible SDFs.

Subject to the weak regularity condition that mean welfare must exist, statistical decision theory is entirely general. It enables comparison of all feasible SDFs. It applies whatever the sampling process may be and whatever information the decision maker may have about the true state of nature and the sampling process. The set of feasible states of nature expresses the assumptions about treatment response and the sampling process that the decision maker finds credible to maintain. This set may be finite-dimensional (parametric) or infinite-dimensional (nonparametric). The maintained assumptions may point-identify or partially identify the true state of nature.

*Statistical Decision Theory and Hypothesis Testing*

Consider again choice between two treatments A and B as in Section 3. In this context, SDFs can be viewed as hypothesis tests where the null hypothesis is that treatment A is better and the alternative is that B is better. However, testing in the Wald theory differs fundamentally from the classical Neyman-Pearson hypothesis testing that is conventionally used in applied statistics and that has been discussed in philosophical consideration of inductive risk.

The Wald theory does not restrict attention to tests that yield a predetermined upper bound on the probability of a Type I error, choosing B when A is better. Nor does it aim to minimize the maximum value of the probability of a Type II error, choosing A when B is better. Wald proposed for binary choice, as elsewhere, evaluation of the performance of SDF $c(\cdot)$ in state s by the expected welfare that it yields across realizations of the sampling process.

Manski (2007, Chapter 12) proved that regret is the probability across repeated samples that an SDF makes an erroneous treatment choice times the magnitude of the loss in welfare when the error occurs. In a state of nature where A is better in terms of mean treatment response, the regret of an SDF is the product of the probability of a type I error (choosing B) and the welfare loss when choosing B. In a state where B is better, regret is the probability of a type II error (choosing A) times the magnitude of the welfare loss when choosing A.

I suggest that using regret to measure inductive risk responds effectively to what Douglas (2000) called for when making treatment decisions. Recall my critique in Section 1 of the use of conventional statistical hypothesis testing to choose treatments. I called attention to the asymmetric treatment of type 1 and type 2 error probabilities and the inattention to magnitudes of welfare losses when errors occur. Evaluation of treatment rules by regret overcomes these problems. Regret considers type I and II error probabilities symmetrically. It measures the magnitudes of the losses that errors produce.

5.3. Concluding Thoughts

Classical welfare economics expresses the values of society via the social welfare function and assumes that scientific knowledge is strong enough to yield a state space with one element, thus enabling optimization of social welfare. As discussed above, statistical decision theory greatly expands the applicability of welfare economics to enable policy comparison in the presence of substantial scientific uncertainty, including both partial identification and statistical imprecision. This expansion of welfare economics copes with Practical Underdetermination without appealing to any auxiliary device, such as 'simplicity," to accept one theory and reject others that are empirically indistinguishable.

I view this as a highly attractive approach to policy analysis. It clearly separates the normative role of society and the scientific role of researchers. Society specifies the SWF and researchers specify the state space. Selection of an SWF and state space are meta-decisions that necessarily must be context specific. I see no conceptual impediment to implementation of the approach. Pragmatic considerations including analytical tractability may be relevant.

A conceptual aspect of welfare-economic planning under uncertainty that some might think unappealing is that, when alternative policies are undominated, one cannot assert that a particular policy is optimal. One can only use some decision criterion—say maximization of subjective expected welfare, maximin, or minimax regret—to assert that a policy choice is 'reasonable.' If one were to accept one theory and reject all others, one might claim optimality. However, the claim would be fragile.